\documentclass{article}
\usepackage[dvips]{graphicx}
\usepackage{fancybox}
\usepackage{amsmath}

\begin{document}
\author{David R. Reichman\footnote{Current address: Department of Chemistry, Columbia University
3000 Broadway, New York, NY 10027
reichman@chem.columbia.edu}\\Patrick
Charbonneau\footnote{pcharbon@fas.harvard.edu}}
\title{Mode-Coupling Theory}
\date{Department of Chemistry and Chemical Biology Harvard University
12 Oxford Street, Cambridge, Massachusetts, 02138} \maketitle

\abstract{In this set of lecture notes we review the mode-coupling
theory of the glass transition from several perspectives. First, we
derive mode-coupling equations for the description of density
fluctuations from microscopic considerations with the use the
Mori-Zwanzig projection operator technique. We also derive schematic
mode-coupling equations of a similar form from a field-theoretic
perspective. We review the successes and failures of mode-coupling
theory, and discuss recent advances in the applications of the
theory.}

\section{Important Phenomenology for MCT}

\begin{figure}
  \includegraphics[width=0.5\columnwidth]{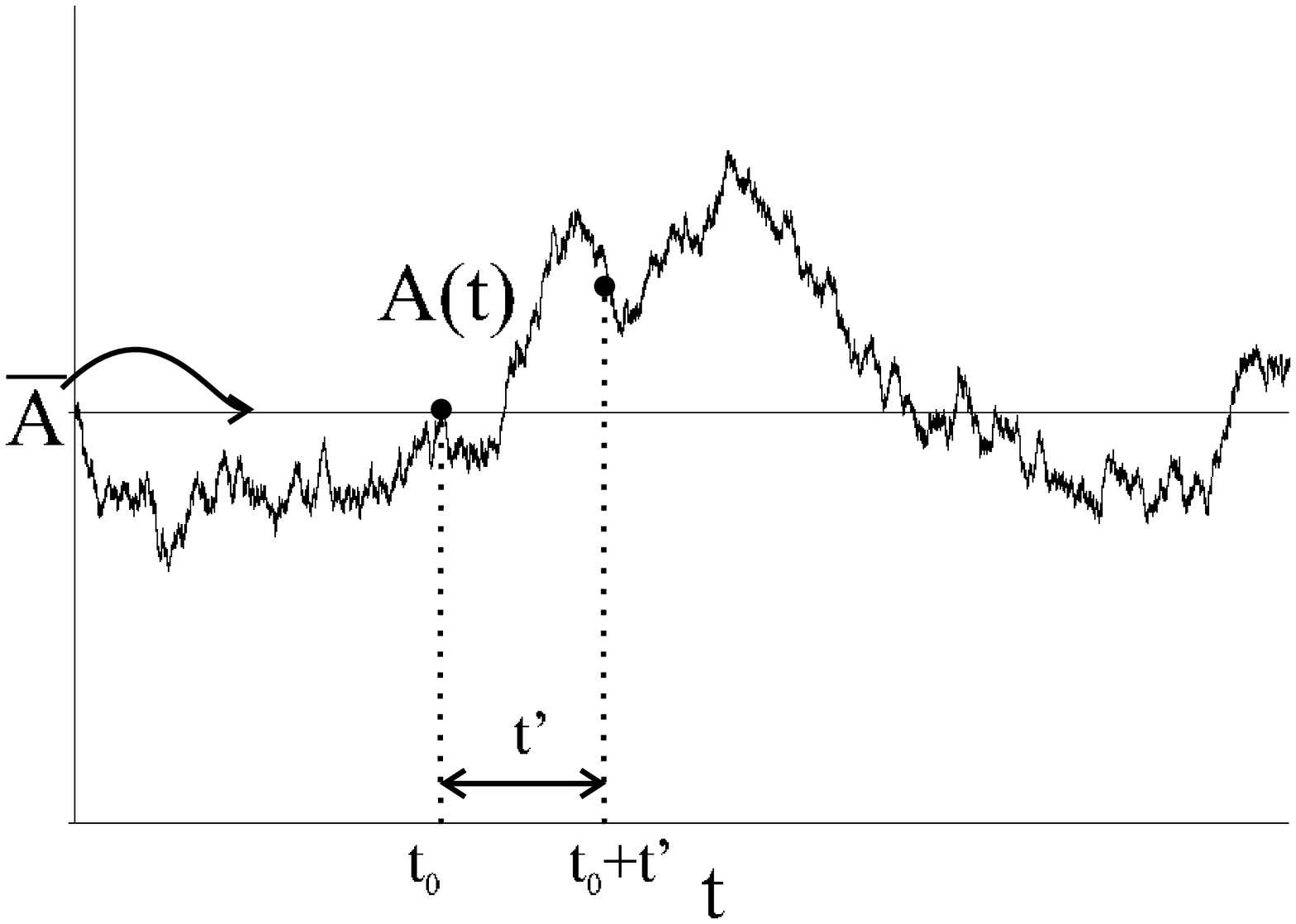}
  \includegraphics[width=0.3\columnwidth]{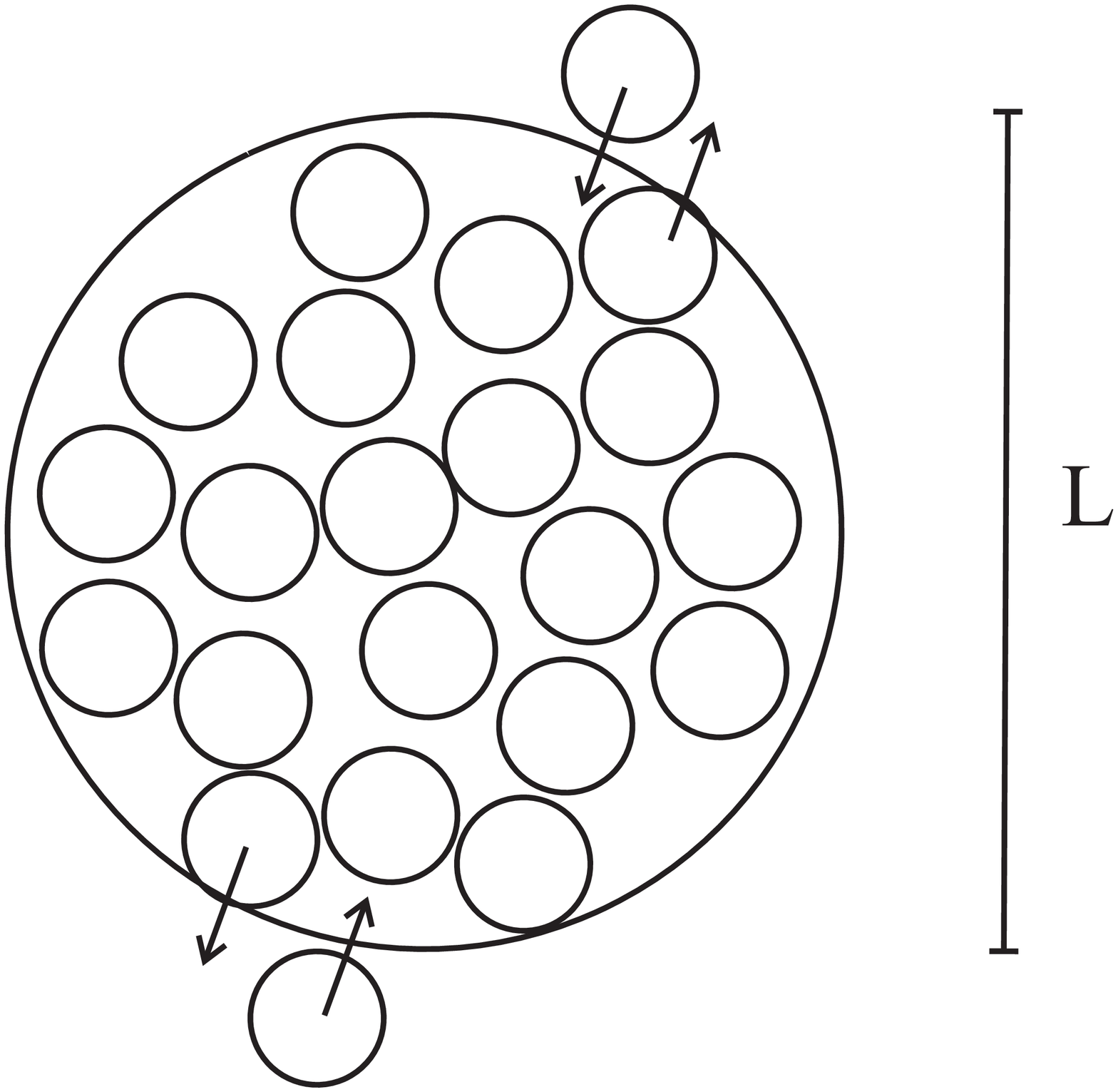}\\
  \caption{Left: time evolution of the instantaneous fluctuations of the quantity $A$.
  The product of fluctuations separated by time $t'$,
  averaged over all $t_0$'s gives the correlation
  function $C(t')$, at equilibrium.
  Right: fluctuations of the density on lengthscale $L\sim 2\pi/k$;
  if $k$ is small, the area of density fluctuations is large.}\label{fig:fluctuations}
\end{figure}
Since our objective will be to sketch a derivation of what we will call Mode-Coupling Theory
(MCT), we will focus our attention on one observable in particular,
namely density fluctuations.  For this we will first define some
of the concepts needed to do so.

We want to calculate a specific time correlation function.  In
general, such a function is expressed as follows,
\begin{eqnarray}
C(t)=\langle A(t) A(0)\rangle\mathrm{.}
\end{eqnarray}
It is an ensemble average of the evolution of the fluctuations of a variable in
time, at equilibrium. As seen in Fig.~\ref{fig:fluctuations}, $A(t)$ fluctuates around its average value in
equilibrium, while $C(t)$ measures the correlation of $A$ at one
time with the value of  $A$ at another time.

The density or particles in a liquid can be one example of $A(t)$,
\begin{eqnarray}
\rho(\mathbf{r},t)= \sum_i
\delta\left(\mathbf{r}-\mathbf{r}_i(t)\right)\mathrm{,}
\end{eqnarray}
which we can Fourier transform,
\begin{eqnarray}
\rho_{\mathbf{k}}(t) &=&\sum_i \int
d\mathbf{r}e^{i
\mathbf{k}\cdot\mathbf{r}}\delta\left(\mathbf{r}-\mathbf{r}_i(t)\right)\nonumber\mathrm{,}\\
&=&\sum_i e^{i \mathbf{k}\cdot\mathbf{r}_i(t)}\mathrm{.}
\end{eqnarray}
In this case the correlation function will be labelled $F(\mathbf{k},t)$,
which is can be expressed as follows.
\begin{eqnarray}\label{eq:kdensity}
F(k,t)=\frac{1}{N}\left\langle\rho_{-\mathbf{k}}(0)\rho_{\mathbf{k}}(t)\right\rangle
=\frac{1}{N}\sum_{ij}\left\langle e^{-i
\mathbf{k}\cdot\mathbf{r}_i(0)}e^{i
\mathbf{k}\cdot\mathbf{r}_j(t)}\right\rangle\mathrm{.}
\end{eqnarray}
Note that we need to have $\sum_i \mathbf{k}_i=0$ (\emph{i.e.}
$-\mathbf{k}+\mathbf{k}=0$!) to conserve momentum, otherwise the
correlation function is equal to zero.

The variables labeled by $\mathbf{k}$ measure density fluctuations
in reciprocal (``$k=|\mathbf{k}|$'') space, which can be thought as
the inverse length. When $k$ is small we are looking at long
lengthscales, as we can see in Fig.~\ref{fig:fluctuations}. When it
is large, we are probing very short scales.

\begin{figure}
  \includegraphics[width=\columnwidth]{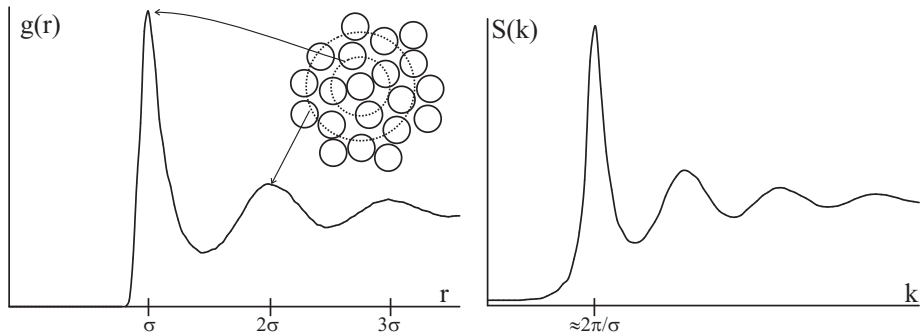}\\
  \caption{Left: radial distribution function $g(r)$ for a simple liquid of size
  $\sigma$. Right: the corresponding structure factor $S(k)$.
  A sample structure is also depicted where the solvation shells are indicated by the dotted lines.
  The exclusion radius can be seen in the absence of amplitude of $g(r)$
  for $r\ll\sigma$.}\label{fig:structure}
\end{figure}
The function $F(k,t)$ is essentially what scattering
experiments measure.  At $t=0$,
\begin{eqnarray}
F(k,t=0)=\frac{1}{N}\left\langle\rho_{-\mathbf{k}}(0)\rho_{\mathbf{k}}(0)\right\rangle
\equiv S(k)\mathrm{,}
\end{eqnarray}
where $S(k)$ is called the static structure factor of the liquid.
Why that name? Consider the radial distribution function of a liquid
$g(r)$. The function $g(r)$ is proportional to the probability that
a particle is a distance $r$ away from a particle at the origin.  In
a dense liquid $g(r)$ shows the structure of the solvation shells as
depicted in Fig.~\ref{fig:structure}. Also, it can be shown that
\cite{balucani:1994,boon:1980,hansen:1986}
\begin{eqnarray}
S(k)=1+ \rho \int d\mathbf{r} e^{-i
\mathbf{k}\cdot\mathbf{r}} g(\mathbf{r})\mathrm{,}
\end{eqnarray}
where $\rho=N/V$ is the density of the system, and thus $S(k)$ is
also indicating something about the liquid structure
\cite{balucani:1994,boon:1980,hansen:1986}. An example for a simple
liquid is depicted in Fig.~\ref{fig:structure}.

\begin{figure}
  \includegraphics[width=\columnwidth]{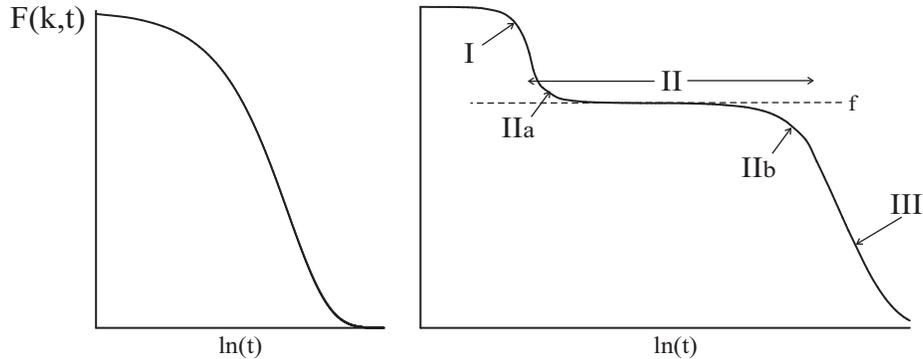}
  \caption{Left: $F(k,t)$ exhibiting exponential $e^{-t/\tau}$ decay for a normal liquid.
  Right: supercooled liquids do not have such a simple decay. The various temporal regimes are described in the text.
  Notice the logarithmic scale.}\label{fig:Fkt}
\end{figure}
But how do we expect $F(k,t)$ to behave? For high temperatures --
above the melting point -- $F(k,t)$ will decay like a single
exponential function in time for for $k \geq 2\pi/sigma$ as plotted
in Fig.~\ref{fig:Fkt}. For supercooled liquids, the situation is
different and a characteristic decay pattern can also be seen in
Fig.~\ref{fig:Fkt}. We observe a multi-step relaxation.
\begin{enumerate}
\item{} At short times decay is coming from free and collisional
events that involve local particle motion.  Consistent with a
short-time expansion, $F(k,t)\sim S(k) - A(k)t^2 + \ldots$ in this
regime \cite{balucani:1994,boon:1980,hansen:1986}.  This will be
true at any temperature. We will not be concerned much with this
part of the decay.

\item{} Intermediate times encompass a period during which
particles appear trapped in cages formed by other particles.  This
regime is the $\beta$-relaxation regime.  The decay to the plateau
(IIa) may be fitted as $f+At^{-a}$ and the decay from the plateau
(IIb) as $f-Bt^b$. Also, the exponents have a scaling
\emph{consistent} with the relationship
\begin{eqnarray}
\frac{\Gamma(1-a)^2}{\Gamma(1-2a)}=\frac{\Gamma(1+b)^2}{\Gamma(1+2b)}\mathrm{.}
\end{eqnarray}

\item{}At long times, in the $\alpha$-relaxation regime, the decay
may be fitted to a stretched exponential law \cite{ediger:1996}
\begin{eqnarray}
F(k,t)\sim e^{-\left(\frac{t}{\tau}\right)^\beta}\mathrm{.}
\end{eqnarray}
with $0<\beta<1$. Do not be confused with the notation. It is the
$\beta$ power that appears in the $\alpha$-relaxation regime! In
general $\beta$ and $\tau$ will be $k$ and temperature dependent.

\end{enumerate}

In a later section, we will return to this kind of phenomenology.
For now, we just make a few superficial remarks about things that
will be covered in more depth by others in these lectures.

The constant $\tau$ that appears in the stretched-exponential decay
law is strongly temperature dependent.  All transports coefficients
--$D$ (diffusion), $\eta$ (viscosity), etc. -- are strongly
temperature dependent as well.  Over a rather wide range of
temperatures, a fit to this temperature dependence may be
\cite{ediger:1996}
\begin{eqnarray} \label{eq:transcoeff}
\eta\sim e^{\frac{E}{T-T_0}}\mathrm{.}
\end{eqnarray}
Clearly, as $T_0$ is approached, relaxation times become so large
that the system cannot stay in equilibrium. Other fitting forms,
some that do not imply a divergence at finite temperatures, may be
used to fit the data as well.

Some systems, hard-spheres for example, are not characterized by
temperature, but by density of packing fraction $\phi = \frac{4}{3}
\pi a^3 \rho$, where $a$ is the particle radius. For such systems,
one may fit with \cite{ediger:1996}
\begin{eqnarray}
\eta\sim e^{\frac{B}{\phi-\phi_c}}\mathrm{,}
\end{eqnarray}
or with other forms.

\section{The Mode-Coupling Theory of density fluctuations}
\label{sec:MCT}

Our strategy will be to derive an \emph{exact} equation of motion
for $F(k,t)$ and then to make approximations that allow us to solve
them \cite{kob:2002,gotze:1999}.  The approximations are
uncontrolled, and we will judge them by their success or failure.

\subsection{Memory functions}

Consider some classical function of phase space variables $A(t)$,
where the time dependence originates from that of the positions
${\mathbf{r}_i}$ and of the momenta ${\mathbf{p}_i}$ for a
$N$-particle system.  We know from Hamilton's equations that
\begin{eqnarray}
\frac{dA(t)}{dt}=\left\{A(t),\mathcal{H}\right\}\equiv
i\mathcal{L}A(t)\mathrm{,}
\end{eqnarray}
where $\{,\}$ is a classical Poisson bracket, which can be expressed
as follows,
\begin{eqnarray}
\left\{A,B\right\}\equiv \sum_i\left(\frac{\partial A}{\partial
\mathbf{r}_i}\cdot\frac{\partial B}{\partial \mathbf{p}_i}-
\frac{\partial A}{\partial \mathbf{p}_i}\cdot\frac{\partial
B}{\partial \mathbf{r}_i}\right)\mathrm{.}
\end{eqnarray}
Also, for liquids of interest, $\mathcal{H}$ is a classical
Hamiltonian with pairwise interactions $\phi(r)$ between the
particles,
\begin{eqnarray}
\mathcal{H}=\sum_i \frac{\mathbf{p}_i^2}{2m} +
\frac{1}{2}\sum_{i,j\neq i}\phi(\mathbf{r}_{ij})\mathrm{.}
\end{eqnarray}
We can thus identify the following,
\begin{eqnarray}
i \mathcal{L} =
\frac{1}{m}\sum_i\left(\mathbf{p}_i\cdot\frac{\partial}{\partial
\mathbf{r}_i}\right)- \sum_{i,j\neq i}\left(\frac{\partial
\phi(\mathbf{r}_{ij})} {\partial \mathbf{r}_i}\cdot
\frac{\partial}{\partial \mathbf{p}_i}\right)\mathrm{.}
\end{eqnarray}
It would be possible to integrate the differential equation to find
$A(t)=e^{i \mathcal{L}t}A(0)$, but this is not useful by itself.

\begin{figure}
\begin{center}
  \includegraphics[width=0.2\columnwidth]{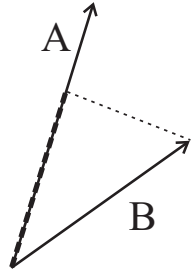}
\end{center}
  \caption{A two-dimensional version of the projection operator $\mathcal{P}B$.
  The quantity $B$ is projected unto the space $A$, which extracts the $A$
  component of $B$ (indicated by a thick dashed line).}\label{fig:projection}
\end{figure}

We also need to define a scalar product of the variables as
\begin{eqnarray}
(A,B)\equiv \langle A^*B\rangle\mathrm{.}
\end{eqnarray}
Now, consider an operator called a \emph{projection
operator}\cite{balucani:1994,boon:1980,hansen:1986,kob:2002,gotze:1999}
$\mathcal{P}$,
\begin{eqnarray}
\mathcal{P}\equiv (A,\ldots)(A,A)^{-1}A\mathrm{.}
\end{eqnarray}
If $A$ is a vector, $(A,A)^{-1}$ is thus the inverse of a matrix. Note
also that $\mathcal{P}^2A=\mathcal{P}A=A$. In geometrical terms, the
projection operator finds the component of some variable $B$ along
the chosen direction $A$, as depicted in Fig.~\ref{fig:projection}.

This is useful as we can extract from an arbitrary $B$ how
much ``character'' of $A$ it has. In particular, the operator $A$ may be a slowly
varying (quasi-hydrodynamic) variable. Consider the density as defined in Eq.~\ref{eq:kdensity},
\begin{eqnarray}
\rho_{\mathbf{k}}(t)=\sum_i e^{i \mathbf{k}\cdot
\mathbf{r}_i(t)}\nonumber\mathrm{,}
\end{eqnarray}
and then
\begin{eqnarray}
\dot{\rho}_{\mathbf{k}}(t)=i \mathbf{k}\cdot \sum_i
\frac{\mathbf{p}_i(t)}{m}e^{i \mathbf{k}\cdot\mathbf{r}_i(t)}
=i\mathbf{k}\cdot
\mathbf{j}_\mathbf{k}(t)=i|\mathbf{k}|j_{\mathbf{k}}^L(t)\mathrm{,}
\end{eqnarray}
where $j^L_{\mathbf{k}}(t)$ is the longitudinal current. If $k$ is
small (large lengthscales), then $\dot{\rho}_{\mathbf{k}}(t)$ is
approximately small. This is what is meant by \emph{slow}. In the
limit $k=0$, then $\dot{\rho}_{k=0} =0$, and the density is strictly
conserved. As Fig.~\ref{fig:fluctuations} indicates, if $k$ is
small, the area of density fluctuations if large, \emph{i.e.} the
rate at which the number of particles fluctuates is small.

We now want to find the exact equation of motion for a correlation
function
\begin{eqnarray}
\frac{d\underline{A}(t)}{dt}&=&e^{i
\mathcal{L}t}\overbrace{\left[\mathcal{P}+(1-\mathcal{P})\right]}^{=1}
i \mathcal{L}\underline{A}\nonumber\mathrm{,}\\
&=&i \underline{\Omega}\cdot \underline{A}(t) + e^{i
\mathcal{L}t}(1-\mathcal{P})i \mathcal{L}\underline{A}\mathrm{,}
\end{eqnarray}
where
\begin{eqnarray}
i \underline{\Omega}=(\underline{A},i
\mathcal{L}\underline{A})\cdot(\underline{A},\underline{A})^{-1}\mathrm{.}
\end{eqnarray}
Now, writing
\begin{eqnarray}
e^{i \mathcal{L}t}=e^{i \mathcal{L}t} \mathcal{O}(t)+
e^{i(1-\mathcal{P})\mathcal{L}t}\mathrm{,}
\end{eqnarray}
$\mathcal{O}(t)$ can be obtained by differentiating both sides of
the equation
\begin{eqnarray}
i \mathcal{L} e^{i \mathcal{L}t}&=&i \mathcal{L} e^{i \mathcal{L}t}
\mathcal{O}(t) + e^{i \mathcal{L}t}\dot{\mathcal{O}}(t) + i
\left[(1-\mathcal{P})\mathcal{L}\right]e^{i(1-\mathcal{P})\mathcal{L}t}
\nonumber\mathrm{,}\\
i \mathcal{L}\left(e^{i \mathcal{L}t} \mathcal{O}(t)+
e^{i(1-\mathcal{P})\mathcal{L}t}\right) &=& i \mathcal{L} e^{i
\mathcal{L}t} \mathcal{O}(t) + e^{i
\mathcal{L}t}\dot{\mathcal{O}}(t) + i
\left[(1-\mathcal{P})\mathcal{L}\right]e^{i(1-\mathcal{P})\mathcal{L}t}
\nonumber\mathrm{,}\\
i \mathcal{L} \mathcal{O}(t)+ e^{i(1-\mathcal{P})\mathcal{L}t}&=&
e^{i \mathcal{L}t}\dot{\mathcal{O}}(t) + i
\left[(1-\mathcal{P})\mathcal{L}\right]e^{i(1-\mathcal{P})\mathcal{L}t}
\nonumber\mathrm{,}\\
e^{i \mathcal{L}t}\dot{\mathcal{O}}(t)&=& i
\mathcal{P}\mathcal{L}e^{i(1-\mathcal{P})\mathcal{L}t}
\nonumber\mathrm{,}\\
\mathcal{O}(t)&=& i\int_0^t d\tau e^{i
\mathcal{L}\tau}\mathcal{P}\mathcal{L}e^{i(1-\mathcal{P})\mathcal{L}\tau}\mathrm{,}
\end{eqnarray}
where the last equality follows from the fact $\mathcal{O}(0)=0$. As
a result, we may write
\begin{eqnarray}\label{eq:projres}
e^{i \mathcal{L}t}i(1-\mathcal{P})\mathcal{L}\underline{A}=
\int_0^td\tau e^{i \mathcal{L}(t-\tau)}i \mathcal{P}\mathcal{L}
f(\tau) + f(t)\mathrm{,}
\end{eqnarray}
where $f(t)$ is called the \emph{fluctuating force}.
\begin{eqnarray}
f(t)\equiv e^{i(1-\mathcal{P})\mathcal{L}t}i
(1-\mathcal{P})\mathcal{L}\underline{A}\mathrm{.}
\end{eqnarray}

What does this mean? The fluctuating force is obtained by taking
the time derivative of $\underline{A}$, using the complimentary
projection operator $(1-\mathcal{P})$ to remove the ``A''
character -- perhaps the \emph{slow} character -- and is then propagated in
the orthogonal  -- \emph{fast} -- space. To put it another way, if
$\mathcal{P}$ removes the slow character from a variable, then the
fluctuating force is the remaining fast force.  We will come back
to this later.

It can be shown that $(\underline{A},f(t))=0$, by noticing that
the definition of $f(t)$ contains the $(1-\mathcal{P})$ factor.
This means that $f(t)$ is orthogonal to $\underline{A}$, in accord
with the discussion above. Noting that
\begin{eqnarray}
i(\underline{A},\mathcal{L}f(t))=i(\mathcal{L}\underline{A},f(t))=
i((1-\mathcal{P})\mathcal{L}\underline{A},f(t))=-(f(0),f(t))\mathrm{,}
\end{eqnarray}
the first term on the RHS of Eq.~\ref{eq:projres} allows the
equation of motion to be rewritten as
\begin{eqnarray}
\frac{d\underline{A}(t)}{dt}= i \underline{\Omega}\cdot
\underline{A}(t)- \int_0^t d\tau \underline{K}(\tau)\cdot
\underline{A}(t-\tau) + f(t)\mathrm{,}
\end{eqnarray}
where we define the \emph{memory function} $\underline{K}(t)$
\begin{eqnarray}
\underline{K}(t)\equiv(f,f(t))\cdot
(\underline{A},\underline{A})^{-1}\mathrm{.}
\end{eqnarray}
This is a fundamental and exact equation for the time dependence
of $\underline{A}(t)$.

Defining the correlation matrix
\begin{eqnarray}
\underline{C}(t)\equiv\langle \underline{A}^*(0)
\underline{A}(t)\rangle = (\underline{A},\underline{A}(t))
\end{eqnarray}
and using the equality $(\underline{A},f(t))=0$, we get
\begin{eqnarray}\label{eq:eqmot}
\frac{d\underline{C}(t)}{dt}= i \underline{\Omega}\cdot
\underline{C}(t)- \int_0^t d\tau \underline{K}(\tau)\cdot
\underline{C}(t-\tau)
\end{eqnarray}
as an exact equation for the matrix of correlation functions
$\underline{C}(t)$ that we will want to compute.  The problem
with computing $\underline{C}(t)$ is embodied in the difficulty of
determining $\underline{K}(t)$.

Now, we want to focus this general framework on density fluctuations
with respect to the bulk density $\rho$, which will allow us to get
an expression for the intermediate scattering function. Consider
\cite{hansen:1986}
\begin{eqnarray}
\underline{A}=
  \begin{bmatrix}
    \delta \rho_\mathbf{q}  \\
    j_{\mathbf{q}}^{L}
  \end{bmatrix}\mathrm{,}
\end{eqnarray}
where
\begin{eqnarray}
\delta \rho_{\mathbf{q}} &=& \sum_i e^{i \mathbf{q}\cdot
\mathbf{r}_i}\nonumber-(2\pi)^3\rho\delta(\mathbf{q})\mathrm{,}\\
j_{\mathbf{q}}^L &=& \frac{1}{m} \sum_i (\mathbf{\hat{q}}\cdot
\mathbf{p}_i)e^{i \mathbf{q}\cdot \mathbf{r}_i}\mathrm{,}
\end{eqnarray}
and therefore
\begin{eqnarray}
\underline{C}(t)&=& \langle
\underline{A}^*\underline{A}(t)\rangle \nonumber\\
&=&
\begin{bmatrix}
  \langle \delta \rho_{\mathbf{-q}} \delta \rho_{\mathbf{q}}(t)\rangle & \langle  \delta \rho_{\mathbf{-q}}j_{\mathbf{q}}^L(t)\rangle \\
  \langle j_{\mathbf{-q}}^L \delta \rho_{\mathbf{q}}(t)\rangle & \langle  j_{\mathbf{-q}}^L j_{\mathbf{q}}^L(t)\rangle
\end{bmatrix}\mathrm{.}
\end{eqnarray}
For the purpose of this demonstration, we will concentrate
on the element in the lower left corner of the matrix, which is in this case
$\frac{N}{i q}\frac{d^2F(q,t)}{dt^2}$. At t=0, the matrix reduces to
\begin{eqnarray}
\underline{C}(0)=
\begin{bmatrix}
  NS(q) & 0 \\
  0 & \frac{Nk_B T}{m}
  \end{bmatrix}\mathrm{.}
\end{eqnarray}
Also,
\begin{eqnarray}
i \underline{\Omega}&=& \langle \underline{A}^*
\underline{\dot{A}}\rangle\cdot\langle
\underline{A}^*\underline{A}\rangle^{-1}\nonumber\mathrm{,}\\
&=&
\begin{bmatrix}
  \langle \delta\rho_{\mathbf{-q}}\delta\dot{\rho}_{\mathbf{q}}\rangle & \langle \delta\rho_{\mathbf{-q}}\frac{d{j}_{\mathbf{q}}^L}{d t}\rangle \\
  \langle j_{\mathbf{-q}}^L\delta\dot{\rho}_{\mathbf{q}}\rangle & \langle \frac{d{j}_{\mathbf{-q}}^L}{d t}\delta\dot{\rho}_{\mathbf{q}}\rangle
\end{bmatrix} \cdot
\langle \underline{A}^*\underline{A}\rangle^{-1}\nonumber\mathrm{,}\\
&=&
\begin{bmatrix}
  0 & i \frac{N qk_B T}{m} \\
  i \frac{Nqk_BT}{m} & 0
\end{bmatrix}\cdot
\begin{bmatrix}
  \frac{1}{NS(q)} & 0 \\
  0 & \frac{m}{Nk_BT}
\end{bmatrix}\nonumber\mathrm{,}\\
&=&
\begin{bmatrix}
  0 & i q \\
  i \frac{qk_BT}{mS(q)} & 0
\end{bmatrix}\mathrm{.}
\end{eqnarray}
To obtain the last two equations, we used integration by parts, the
property that the correlation of an observable and its derivative is
always zero, and the statistical thermodynamics result that follows:
\begin{eqnarray}\label{eq:equipartition}
\langle j_{\mathbf{-q}}^L\delta\dot{\rho}_{\mathbf{q}}\rangle&=&\frac{i}{m^2}
\sum_{i,j} \left\langle(\mathbf{\hat{q}}\cdot \mathbf{p}_i) e^{-i
\mathbf{q}\cdot \mathbf{r}_i} (\mathbf{q}\cdot
\mathbf{p}_j)e^{i \mathbf{q}\cdot \mathbf{r}_j}\right\rangle\nonumber\mathrm{,}\\
&=& \frac{i q}{m}\sum_i \left\langle m v_i^2\right\rangle = i\frac{N
q k_BT}{m}\mathrm{.}
\end{eqnarray}

The random force $f(0)$ is expressed as
\begin{eqnarray}
f(0) &=&(1-\mathcal{P})\underline{\dot{A}}\nonumber\mathrm{,}\\
&=&
\begin{bmatrix}
  \delta \dot{\rho}_{\mathbf{q}} \\
  \frac{dj_{\mathbf{q}}^L}{dt}
\end{bmatrix}
-
\begin{bmatrix}
  0 & i q \\
  i\frac{qk_BT}{mS(q)} & 0
\end{bmatrix}
\cdot
\begin{bmatrix}
  \delta \rho_{\mathbf{q}} \\
  j_{\mathbf{q}}^L
\end{bmatrix}\nonumber\mathrm{,}\\
&=&
\begin{bmatrix}
  0 \\
  \frac{dj_{\mathbf{q}}^L}{dt}-i\frac{qk_BT}{mS(q)}\delta \rho_{\mathbf{q}}
\end{bmatrix}
\equiv
\begin{bmatrix}
  0 \\
  R_{\mathbf{q}}
\end{bmatrix}\mathrm{.}
\end{eqnarray}
We will now look at the equation of motion term by term. First,
\begin{eqnarray}
\frac{d\underline{C}(t)}{dt} =
\begin{pmatrix}
  \frac{d}{dt}\langle \delta \rho_{\mathbf{-q}} \delta \rho_{\mathbf{q}}(t)\rangle & \frac{d}{dt}\langle \delta \rho_{\mathbf{-q}}j_{\mathbf{q}}^L(t)\rangle \\
  \frac{d}{dt}\langle j_{\mathbf{-q}}^L \delta \rho_{\mathbf{q}}(t)\rangle & \frac{d}{dt}\langle j_{\mathbf{-q}}^L j_{\mathbf{q}}^L(t)\rangle
\end{pmatrix}\mathrm{.}
\end{eqnarray}
Note that the lower left corner term equals $\frac{N}{i
q}\frac{d^2F(q,t)}{dt^2}$. Second,
\begin{eqnarray}
i \underline{\Omega}\cdot\underline{C}(t)&=&
\begin{bmatrix}
  0 & i q \\
  i\frac{qk_BT}{mS(q)} & 0
\end{bmatrix} \cdot
\begin{bmatrix}
  \langle \delta\rho_{\mathbf{-q}}\delta\rho_{\mathbf{q}}(t)\rangle & \langle \delta\rho_{\mathbf{-q}} j_{\mathbf{q}}^L(t)\rangle \\
  \langle j_{\mathbf{-q}}^L \delta \rho_{\mathbf{q}}(t)\rangle & \langle j_{\mathbf{-q}}^L j_{\mathbf{q}}^L(t)\rangle
\end{bmatrix}\mathrm{.}
\end{eqnarray}
Notice the lower left corner term is $-\frac{q N
k_BT}{imS(q)}F(q,t)$ this time. Lastly, the memory matrix is
\begin{eqnarray}
\underline{K}(t)&=&\left\langle
\begin{bmatrix}
  0 \\
  R_{\mathbf{q}}^*
\end{bmatrix}\cdot
\begin{bmatrix}
  0 & R_{\mathbf{q}}(t)
\end{bmatrix}
\right\rangle \cdot \langle
\underline{A}^* \underline{A}\rangle^{-1} \nonumber\mathrm{,}\\
&=&
\begin{bmatrix}
  0 & 0 \\
  0 & \langle R_{\mathbf{-q}}R_{\mathbf{q}}(t)\rangle
\end{bmatrix}\cdot
\begin{bmatrix}
  \frac{1}{NS(q)} & 0 \\
  0 & \frac{m}{Nk_BT}
\end{bmatrix}\nonumber\mathrm{,}\\
&=&
\begin{bmatrix}
  0 & 0 \\
  0 & \frac{m\langle R_{\mathbf{-q}}R_{\mathbf{q}}(t)\rangle}{Nk_BT}
\end{bmatrix}\mathrm{.}
\end{eqnarray}
Concentrating on the lower left corner, using the equation of motion
from Eq.~\ref{eq:eqmot}, we find
\cite{balucani:1994,boon:1980,hansen:1986,kob:2002}
\begin{eqnarray}\label{eq:exactmemory}
\frac{d^2F(q,t)}{dt^2}+\frac{q^2k_BT}{mS(q)}F(q,t) +
\frac{m}{Nk_BT}\int_0^t d\tau \langle R_{\mathbf{-q}}R_{\mathbf{q}}(\tau)\rangle
\frac{d}{dt}F(q,t-\tau) = 0\mathrm{.}
\end{eqnarray}

This equation is exact, but impossible to solve.  To make
approximations, we will look at $\langle R_{\mathbf{-q}}R_{\mathbf{q}}(t)\rangle$
using some intuition. Recall that
\begin{eqnarray}
R_{\mathbf{q}}=\frac{dj_{\mathbf{q}}^L}{dt} - i\frac{qk_BT}{mS(q)} \delta \rho_{\mathbf{q}}
\end{eqnarray}
and that
\begin{eqnarray}\label{eq:djdt}
\frac{dj_{\mathbf{q}}^L}{dt}&=&\frac{d}{dt}\left\{\frac{1}{m}\sum_i(\mathbf{\hat{q}}\cdot\mathbf{p}_i)
e^{i \mathbf{q}\cdot\mathbf{r}_i}\right\}\nonumber\mathrm{,}\\
&=&
\frac{1}{m}\sum_i\left(\mathbf{\hat{q}}\cdot\frac{d\mathbf{p}_i}{dt}\right)
e^{i \mathbf{q}\cdot\mathbf{r}_i} +
\frac{i}{m^2}\sum_i\left(\mathbf{\hat{q}}\cdot\mathbf{p}_i\right)^2
e^{i \mathbf{q}\cdot\mathbf{r}_i}\mathrm{.}
\end{eqnarray}
Also, note that in this last equation $\frac{d\mathbf{p}_i}{dt}$
is a force and therefore
\begin{eqnarray}
\frac{d\mathbf{p}_i}{dt}\sim-\sum_{i\neq j}\nabla\phi(|\mathbf{r}_i
- \mathbf{r}_j|)=\sum_\mathbf{k} i \mathbf{k}\phi_{\mathbf{k}}
\delta\rho_{\mathbf{k}} \delta\rho_{-\mathbf{k}}\mathrm{,}
\end{eqnarray}
where we made the momentum-space transformation $\phi_{\mathbf{k}} =
\int d\mathbf{r}e^{i\mathbf{k}\cdot\mathbf{r}}\phi(\mathbf{r})$. We
then see that hidden in the fluctuating force is a pair of
densities.  This illustrates an important point: at first, we may
have suspected that the fluctuating force is a \emph{fast} variable
and that it decays on a short timescale because we removed the slow
modes $\delta \rho_{\mathbf{k}}$ from it -- but we see that it
contains, at leading order, a slow character (at least if
$\delta\rho_{\mathbf{k}}$ is slow) from the product of slow modes
$\delta\rho_{\mathbf{k}}\delta\rho_{\mathbf{-k}}$! Overall the time
derivative of the current has the symmetry of $\delta
\rho_{\mathbf{k}} \delta\rho_{\mathbf{q}-\mathbf{k}}$, where the
$\mathbf{q}$ factor comes from $\sum_i
e^{i\mathbf{q}\cdot\mathbf{r}_i}$ that multiplies the force in
Eq.~\ref{eq:djdt}.

We will now approximate $\langle
R_{\mathbf{-q}}e^{i\mathcal{Q}\mathcal{L}t}R_{\mathbf{q}}\rangle$. As a convention,
note that $\mathcal{Q}\equiv 1-\mathcal{P}$.
\begin{enumerate}
\item{}Replace
$e^{i\mathcal{Q}\mathcal{L}t}\rightarrow\mathcal{P}_2e^{i
\mathcal{L}t}\mathcal{P}_2$, where we define the new projection operator
\begin{eqnarray}
\mathcal{P}_2\equiv\sum_{\mathbf{k_1},\mathbf{k_2},\mathbf{k_3},\mathbf{k_4}}
A_{\mathbf{k_1},\mathbf{k_2}}\langle
A^*_{\mathbf{k_3},\mathbf{k_4}}\ldots\rangle \langle
A^*_{\mathbf{k_1},\mathbf{k_2}}
A_{\mathbf{k_3},\mathbf{k_4}}\rangle^{-1}
\end{eqnarray}
and where $A_{\mathbf{k_1},\mathbf{k_2}}=\delta \rho_\mathbf{k_1}
\delta \rho_{\mathbf{k_2}}$. The $\mathcal{P}_2$ operator simply
projects $R_{\mathbf{q}}$ onto its dominant slow product mode. We
neglect the $\mathcal{Q}$ in the exponent, simply because it is hard
to compute anything keeping it there.  However, to
$\mathcal{O}(q^2)$, it may be shown that this neglect is not
consequential.

\item Factorize four-point density terms into products of
two-point ones.
\end{enumerate}

Using this algorithm, we  get
\begin{eqnarray}
\mathcal{P}_2 R_{\mathbf{q}}=\sum_{\mathbf{k_1},\mathbf{k_2}}
V_{\mathbf{q}}(\mathbf{k_1},\mathbf{k_2})\delta \rho_{\mathbf{k_1}}
\delta \rho_{\mathbf{k_2}}\mathrm{,}
\end{eqnarray}
where
\begin{eqnarray}
V_{\mathbf{q}}(\mathbf{k_1},\mathbf{k_2})&\equiv&
\sum_{\mathbf{k_3},\mathbf{k_4}}\left\langle
\delta\rho_{\mathbf{k_1}}\delta
\rho_{\mathbf{k_2}}R_{\mathbf{q}}\right\rangle \cdot\left\langle
\delta\rho_{\mathbf{k_1}}\delta \rho_{\mathbf{k_2}}\delta
\rho_{\mathbf{k_3}}\delta \rho_{\mathbf{k_4}}\right\rangle^{-1}
\end{eqnarray}
The denominator has a product of four density variables that will be
factorized into products of two structure factors.  The numerator
has terms like:
\begin{eqnarray} \label{eq:1sttermnumerator}
\left\langle\delta \rho_{-\mathbf{k}}\delta
\rho_{\mathbf{k}-\mathbf{q}}\frac{dj_{\mathbf{q}}^L}{dt}\right\rangle=-\langle
\delta\dot{\rho}_{-\mathbf{k}}\delta
\rho_{\mathbf{k}-\mathbf{q}}j_{\mathbf{q}}^L\rangle -\langle
\delta\rho_{-\mathbf{k}}\delta
\dot{\rho}_{\mathbf{k}-\mathbf{q}}j_{\mathbf{q}}^L\rangle\mathrm{,}
\end{eqnarray}
where the result was obtained by integration by parts, and
\begin{eqnarray}\label{eq:2ndtermnumerator}
-\frac{i q k_BT}{mS(q)}\langle \delta
\rho_{-\mathbf{k}}\delta\rho_{\mathbf{k}-\mathbf{q}}\delta \rho_{\mathbf{q}}\rangle\mathrm{.}
\end{eqnarray}
Let's calculate one of the terms in Eq.~\ref{eq:1sttermnumerator},
\begin{eqnarray}
-\langle \delta\dot{\rho}_{-\mathbf{k}}\delta \rho_{\mathbf{k}-\mathbf{q}}j_{\mathbf{q}}^L\rangle &=& i
\left\langle \sum_i (\mathbf{k}\cdot\mathbf{p}_i)e^{-i
\mathbf{k}\cdot\mathbf{r}_i}\sum_j e^{i
(\mathbf{k}-\mathbf{q})\cdot\mathbf{r}_i}\sum_l(\mathbf{\hat{q}}\cdot\mathbf{p}_l)
e^{i
\mathbf{q}\cdot\mathbf{r}_l}\right\rangle \nonumber\mathrm{,}\\
&=& i \sum_{ij}\left\langle
e^{i(\mathbf{k}-\mathbf{q})\cdot\mathbf{r}_j}
e^{i(\mathbf{q}-\mathbf{k})\cdot\mathbf{r}_i}\right\rangle
\frac{k_BT}{m}(\mathbf{k}\cdot\mathbf{\hat{q}})\nonumber\mathrm{,}\\
&=& i (\mathbf{k}\cdot\mathbf{\hat{q}})
\frac{k_BT}{m}NS(|\mathbf{k}-\mathbf{q}|)\mathrm{,}
\end{eqnarray}
where we used the result of Eq.~\ref{eq:equipartition},
to complete the calculation. The
other term similarly gives
\begin{eqnarray}
-\langle
\delta\rho_{-\mathbf{k}}\delta \dot{\rho}_{\mathbf{k}-\mathbf{q}}j_{\mathbf{q}}^L\rangle =
i\left(\mathbf{\hat{q}}\cdot(\mathbf{q}-\mathbf{k})\right)\frac{k_BT}{m}NS(k)\mathrm{.}
\end{eqnarray}
The term in Eq.~\ref{eq:2ndtermnumerator} is hard to compute
directly, but within the \emph{convolution approximation}, it can be
reduced as follows \cite{balucani:1994,kob:2002}:
\begin{eqnarray}
\langle \delta \rho_{-\mathbf{k}}\delta\rho_{\mathbf{k}-\mathbf{q}}\delta \rho_{\mathbf{q}}\rangle \approx
NS(k)S(q)S(|\mathbf{k}-\mathbf{q}|)\mathrm{.}
\end{eqnarray}
After treating all static density fluctuations within the Gaussian
(and convolution) approximations, we find that the vertex
$V_{\mathbf{q}}(\mathbf{k_1},\mathbf{k_2})$ can be expressed as a
function of only two wavevectors. As a consequence of translational
invariance, we would be left only with terms involving the
difference of wavevectors
$\mathbf{k}\equiv\mathbf{k_1}-\mathbf{k_2}$, which allows us to
write
$V_{\mathbf{q}}(\mathbf{k_1},\mathbf{k_2})=V_{\mathbf{k},\mathbf{q}-\mathbf{k}}$.
Also note that the summation is now only over $\mathbf{k}$.
Combining all terms gives
\begin{eqnarray}
V_{\mathbf{k},\mathbf{q}-\mathbf{k}}&=& \frac{i
k_BT}{2mN}\left\{\frac{(\mathbf{\hat{q}}\cdot\mathbf{k})}{S(k)} +
\frac{\mathbf{\hat{q}}\cdot(\mathbf{q}-\mathbf{k})}{S(|\mathbf{k}-\mathbf{q}|)}
-(\mathbf{q}\cdot\mathbf{\hat{q}})\right\}
\nonumber\mathrm{,}\\
&=& \frac{i \rho k_BT}{2mN}\left\{(\mathbf{\hat{q}}\cdot\mathbf{k})c(k)+
\mathbf{\hat{q}}\cdot(\mathbf{q}-\mathbf{k})c(|\mathbf{k}-\mathbf{q}|)\right\}\mathrm{,}
\end{eqnarray}
where we have rewritten the result using the direct correlation function $c(k)\equiv
\frac{1}{\rho}\left(1-\frac{1}{S(k)}\right)$. So, piecing this together,
\begin{eqnarray}
\left\langle (R_{\mathbf{q}}\mathcal{P}_2)^*
(\mathcal{P}_2R_{\mathbf{q}}(t))\right\rangle &\simeq&
\sum_{\mathbf{k},\mathbf{k'}}\left|V_{\mathbf{k},\mathbf{q}-\mathbf{k}}^*
V_{\mathbf{k'},\mathbf{q}-\mathbf{k'}}\right| \left\langle\delta
\rho_{-\mathbf{k'}}\delta\rho_{\mathbf{k'}-\mathbf{q}}\delta
\rho_{\mathbf{k}}(t)\delta\rho_{\mathbf{q}-\mathbf{k}}(t)\right\rangle \nonumber\mathrm{,}\\
&\simeq&
\sum_{\mathbf{k},\mathbf{k'}}\left|V_{\mathbf{k},\mathbf{q}-\mathbf{k}}^*
V_{\mathbf{k'},\mathbf{q}-\mathbf{k'}}\right|
N^2 F(k,t)F(|\mathbf{q}-\mathbf{k}|,t)(\delta_{\mathbf{k},\mathbf{k'}}+\delta_{\mathbf{k'}-\mathbf{q},\mathbf{k}})\nonumber\mathrm{,}\\
&=&
\frac{\rho^2(k_BT)^2}{2m^2}\sum_{\mathbf{k}}\left|\tilde{V}_{\mathbf{q}-\mathbf{k},\mathbf{k}}\right|^2F(k,t)
F(|\mathbf{q}-\mathbf{k}|,t)\mathrm{,}
\end{eqnarray}
where we used Wick's factorization and where
\begin{eqnarray}
\tilde{V}_{\mathbf{q}-\mathbf{k},\mathbf{k}}\equiv\left\{(\mathbf{\hat{q}}\cdot\mathbf{k})c(k)+
\mathbf{\hat{q}}\cdot(\mathbf{q}-\mathbf{k})c(|\mathbf{q}-\mathbf{k}|)\right\}\mathrm{.}
\end{eqnarray}
We need to convert the discrete sum to the continuous integral,
$\sum_\mathbf{k} \rightarrow \frac{V}{(2\pi)^3}\int d\mathbf{k}$, and
multiply by the $\frac{m}{Nk_BT}$ prefactor as obtained in
Eq.~\ref{eq:exactmemory}, to get the final MCT equation.

\noindent\fbox{
\begin{Beqnarray}\label{eq:MCTeq}
0 &=& \frac{d^2F(q,t)}{dt^2}+ \frac{q^2k_BT}{mS(q)}F(q,t)+\int_0^t
d\tau K(q,t-\tau)\frac{\partial F(q,\tau)}{\partial \tau}\\
\mathrm{with}\nonumber\\
K(q,t)
&=&\frac{\rho k_BT}{16\pi^3 m}\int d\mathbf{k}|\tilde{V}_{\mathbf{q}-\mathbf{k},\mathbf{k}}|^2
F(k,t)F(|\mathbf{k}-\mathbf{q}|,t)
\end{Beqnarray}
}

\subsection{Some Properties of the Solution(s) of the MCT
Equation}

\subsubsection{Schematic MCT}
\begin{figure}
  \includegraphics[width=\columnwidth]{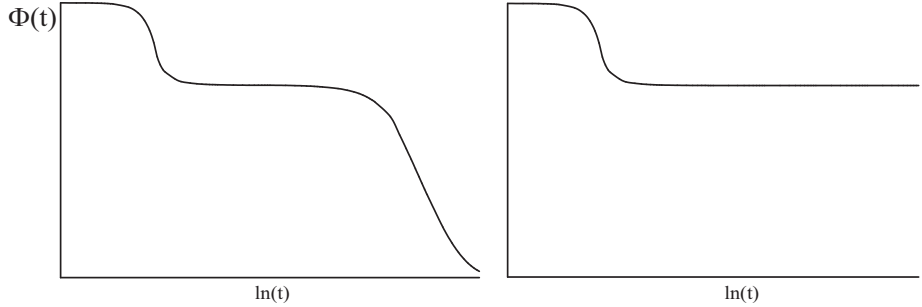}
  \caption{Left: $\Phi(t)$ decay in the ergodic (supercooled) case. The
  correlation vanishes on a finite timescale.
  Right: in the non-ergodic (glassy) case, that same function remains
  finite even for infinite times.}\label{fig:phit}
\end{figure}
Via the approximation discussed by Bengtzelius \emph{et al.} we can
reduce our MCT equation to a schematic form,
\begin{eqnarray}\label{eq:schematic}
\frac{\partial^2 \Phi(t)}{\partial t^2} + \Omega_0^2 \Phi(t) +
\lambda \int_0^t d\tau \Phi^2(t-\tau)\frac{\partial
\Phi(\tau)}{\partial \tau} =0\mathrm{,}
\end{eqnarray}
where $\Phi(t)\sim F(k,t)$, before we neglect the coupling
wavevectors. The solutions of Eq.~\ref{eq:schematic} have been
discussed by Leutheusser~\cite{leutheusser:1984} and Bengtzelius
\emph{et al.}~\cite{bengtzelius:1984}. The most striking feature of
this equation, and of the full MCT equation from which we
``derived'' it, is that there is a transition to a completely
non-ergodic phase for particular $\Omega_0^2$ and $\lambda$ (or $T$
and $\rho$ for the \emph{real} MCT equation). The two cases are
depicted in Fig.~\ref{fig:phit}.

The transition from ergodic to non-ergodic at a sharp,
well-defined set of parameters
may be interpreted as the transition from a liquid to a
solid. The fact that correlations do not decay as $t\rightarrow
\infty$ is indicative of this. However, no information of an
ordered state was used or imposed.  Thus, the solid could only be
a disordered one, \emph{i.e. a glass}.

\subsubsection{Solutions of Full MCT Equations}

If we denote by $T_c$ the temperature where MCT predicts a glass
transition, the relaxation times $\tau$ scales as
\cite{kob:2002,gotze:1999,leutheusser:1984,bengtzelius:1984}
\begin{eqnarray}\label{eq:relaxtime}
\tau(q,T) \sim A_q (T-T_c)^{-\gamma}\mathrm{,}
\end{eqnarray}
which means that when $T\rightarrow T_c$, $\tau$ diverges as a power
law with a universal exponent $\gamma$. This form may fit data, but
only over a limited temperature range.

Also, the decay in the $\beta$-relaxation regime is indeed given by
\cite{kob:2002,gotze:1999,leutheusser:1984,bengtzelius:1984}
\begin{center}
\begin{tabular}{lc}
  early $\beta$: & $f+At^{-a}$,\\
  late $\beta$: & $f-Bt^{b}$, \\
\end{tabular}
\end{center}
with
\begin{eqnarray}
\frac{\Gamma(1-a)^2}{\Gamma(1-2a)}=\frac{\Gamma(1+b)^2}{\Gamma(1+2b)}\mathrm{,}
\end{eqnarray}
at least for $T$ very close to $T_c$, \emph{i.e.} when
$\frac{T-T_c}{T_c}\ll 1$. This is a great triumph of the MCT
equations and is \emph{fully consistent} with simulations and
experiments \cite{kob:2002,gotze:1999}. Furthermore, the power
$\gamma$ is related to the exponents $a$ and $b$ as
\begin{eqnarray}\label{eq:gammaexp}
\gamma=\frac{1}{2a}+\frac{1}{2b}\mathrm{.}
\end{eqnarray}

For the $\alpha$-relaxation regime, an approximate solution of the full
MCT equations is indeed approximately given by
\begin{eqnarray}
F(k,t) \sim e^{-\left(\frac{t}{\tau_k}\right)^{\beta_k}}\mathrm{,}
\end{eqnarray}
and this stretched exponential function describes well experiments
and simulations. The schematic equation, however, only exhibits exponential decay.

More generally, MCT predicts that for a correlator at temperature
$T$ a \emph{time-temperature} superposition holds
\cite{kob:2002,gotze:1999}:
\begin{eqnarray}
C(t,T)=\hat{C}(t/\tau(T))\mathrm{,}
\end{eqnarray}
where $C(t,T)$ is a correlation function, $\hat{C}$ is some
master function and $\tau(T)$ is the $\alpha$-relaxation time.
This is also generally consistent with experiments and simulations.

\subsubsection{Redux: An Assessment of the Successes and Failures
of MCT}
Even a scientist who is opposed to the spirit and approximations
that go into MCT ought to be impressed by its success, where it
succeeds. Furthermore, it is essentially the \emph{only}
first-principle theory of glassy liquids.  Namely, from the
structure of the liquid alone ($S(k)$, the structure factor) a
detailed set of dynamical predictions emerge.  We will now spell
out where MCT works and where (we think) it does not.

\paragraph{Successes}
\begin{enumerate}
\item{} MCT makes some remarkable predictions that \emph{are}
correct.  For example, the remarkable scaling properties in the
$\beta$-relaxation regime that are predicted are essentially correct
and so is the time-temperature superposition in the
$\alpha$-relaxation regime. This is similarly accurate for other
predictions that we will not discuss here
\cite{kob:2002,gotze:1999}.
\item{} MCT has predicted novel relaxation patterns correctly.
One recent striking example is the behavior of colloidal suspensions
with induced short-ranged attractions\cite{dawson:2001}.  Here, MCT
has predicted that adding attractions may melt the glass from
hard-spheres, and that for certain parameters, logarithmic
relaxation may be observed. Both predictions have been confirmed
again by computer simulations and experiments.
\item{} Although we will not discuss it here, there exist models
with quenched disorder (spin-glass models) for which the schematic
MCT is \emph{exact} \cite{bouchaud:1996,bouchaud:1997}. These models
make connections between MCT and energy landscape theories possible,
as well as extensions of the MCT approach to situations that are
out-of-equilibrium (aging). This has been very fruitful and has led
to new insights into glassy systems.
\end{enumerate}

\paragraph{Failures}
\begin{figure}
  \includegraphics[width=0.8\columnwidth]{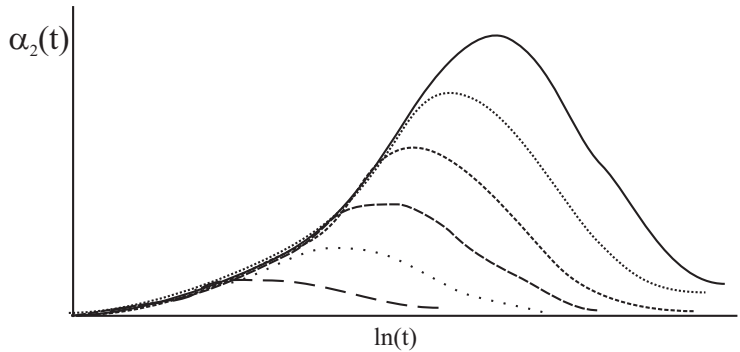}
  \caption{The non-Gaussian parameter as depicted here is
  one of the features that MCT cannot reproduce accurately.
  The curves increase in magnitude and spread with
  decreasing temperature (from left to right).}\label{fig:alpha}
\end{figure}
\begin{enumerate}
\item{} The best-known failure of MCT is that it predicts a sharp
glass transition at a temperature $T_c$, but $T_c>T_g$.  This
means that MCT predicts kinetic arrest to a non-ergodic phase at
temperatures where the system is still ergodic and liquid.
\item{} MCT also predicts power-law divergence of transport coefficients and
relaxation times as in Eq.~\ref{eq:relaxtime}, but this is less
accurate over a wide range of temperatures than the temperature
dependence of transport coefficients given in
Eq.~\ref{eq:transcoeff} \cite{ediger:1996}.  It is a reasonable
fitting form over several decades of relaxation time in mildly
supercooled liquids. In addition, the parameters $\beta$, $a$, $b$,
$\ldots$ are predicted to be constant in MCT, at least for
temperatures for $\frac{T-T_c}{T_c}\ll 1$. But in actuality, they
are mildly temperature dependent.  One should be careful, however,
not to take asymptotic predictions of MCT and apply them to cases
where $\frac{T-T_c}{T_c}$ is not small \cite{kob:2002}.
\item{} Another failure of MCT is in the prediction of certain
indicators of collective relaxation.  In general, timescales and
lengthscales of such heterogeneous motion can be probed by
multi-point correlations.  A simple, non-multi-point function that
seems to correlate crudely with the timescale of such motion is the
non-Gaussian parameter
\begin{eqnarray}
\alpha_2(t)=\frac{3\langle r^4(t)\rangle}{5\langle
r_2(t)\rangle^2}-1\mathrm{,}
\end{eqnarray}
where
\begin{eqnarray}
\langle r^2(t)\rangle = \left\langle [r(t)-r(0)]^2\right\rangle
\end{eqnarray}
for a tagged particle and similarly for the other term. Usually, the
behavior of $\langle \alpha^2(t)\rangle$ is similar to what is
depicted in Fig.~\ref{fig:alpha}. However, MCT predictions of
$\alpha_2(t)$ are quite inaccurate. In general, MCT fails to
accurately depict properties such as the non-Gaussian parameter and
the breakdown of the Stokes-Einstein relation.
\end{enumerate}
One may take from this last result that MCT is not capable of saying
anything about dynamically heterogeneous motion in supercooled
liquids \cite{donati:1999}, but perhaps this statement is too
strong.  We will explore this further in the next section.

\subsection{Field-Theoretic Description}
Before finishing this section, we provide a sketch of the
field-theoretic approach to schematic MCT
\cite{bouchaud:1996,bouchaud:1997}. In some sense, the memory
function approach can be thought of as arising from  coupled
Langevin equations for the modes $\delta \rho_{\mathbf{q}}$ and
$j_{\mathbf{q}}^L$.  For example,
\begin{eqnarray}
\delta\dot{\rho}_{\mathbf{q}} &=& i q j_{\mathbf{q}}^L(t)\mathrm{,}\\
\frac{\partial}{\partial t} j_{\mathbf{q}}^L(t) &=& -\frac{i q
k_BT}{mS(q)} \delta \rho_{\mathbf{q}}(t) - \frac{k_BT}{m}\int
d\mathbf{k} \underbrace{(i \mathbf{\hat{q}}\cdot \mathbf{k})
c(k)\delta
\rho_{\mathbf{q}-\mathbf{k}}(t)\delta\rho_{\mathbf{k}}(t)}_{\text{from
the fluctuating
force}}\nonumber\\
&& -\underbrace{\frac{\zeta_0}{m}j_{\mathbf{q}}^L(t)}_{\text{viscosity term}}
+\underbrace{\eta_{\mathbf{q}}(t)}_{\text{noise term}}\mathrm{.}
\end{eqnarray}
As a toy model for this (forgetting vector labels and
wavevectors), we get
\begin{eqnarray}\label{eq:toyLangevin}
\dot{\phi}(t)=-\mu(t)\phi(t) - \frac{g}{2}\phi(t)^2 + \eta(t)
\end{eqnarray}
with $\langle \eta(t)\eta(0)\rangle = 2T\delta(t)$. We also define
\begin{eqnarray}
G_0(t,t') \equiv e^{-\int_{t'}^t d\tilde{t} \mu(\tilde{t})}
\end{eqnarray}
and
\begin{eqnarray}
(G\bigotimes f)(t)\equiv \int_0^t dt' G_0(t,t') f(t')\mathrm{,}
\end{eqnarray}
where $G_0$ is the \emph{bare} response function, or propagator.  It must be zero if
$t<t'$. The solution for $\phi(t)$, with $\phi(0)=0$ is, in
graphical terms,
\begin{eqnarray}
\phi(t) =
\end{eqnarray}
where an arrow represents $G_{0}$, an $\times$ represents the noise,
and a factor of $g/2$ is associated with each branching point. These
terms are simply obtained as a solution from integrating
Eq.~\ref{eq:toyLangevin}. In a more compact notation,
\begin{eqnarray}
\phi(t)=G_0 \bigotimes \eta - \frac{g}{2} G_0\bigotimes
\left\{G_0\bigotimes \eta\cdot G_0\bigotimes\eta\right\} + \ldots\mathrm{,}
\end{eqnarray}
where ``$\cdot$'' is a simple product.

We define two kinds of functions,
\begin{eqnarray}
C(t,t')\equiv\langle \phi(t)\phi(t')\rangle
\end{eqnarray}
and
\begin{eqnarray}
G(t,t')\equiv\left\langle \frac{\partial
\phi(t)}{\partial\eta(t')}\right\rangle=\frac{1}{2T}\langle\phi(t)\eta(t')\rangle\mathrm{,}
\end{eqnarray}
where the last equality is true for Gaussian noise.  Again,
$C(t,t')$ can be defined on the entire $(t,t')$ plane but $G(t,t')$
only for $t>t'$. Let's construct a series for the two functions. The
zeroth-order contribution to $C(t,t')$ is given as follows:
\begin{eqnarray}
\left\langle
\right\rangle
&=&
\left\langle \int_0^t dt_1 G_0(t,t_1) f(t_1)
\int_0^{t'} dt_2 G_0(t',t_2)f(t_2)\right\rangle \nonumber\mathrm{,}\\
&=&
\int_0^t \int_0^{t'} dt_1 dt_2 G_0(t,t_1)G_0(t',t_2) 2T \delta(t_1-t_2)\mathrm{.}
\end{eqnarray}
The bracket average implies connecting the $\times$-vertices to
form diagrams with none of these left.  Diagrams that do not pair up all
noise vertices (\emph{i.e.} those with an odd number of $\times$-vertices) average to zero.

Going beyond zeroth-order, we get the following:
\begin{eqnarray}\label{eq:Cdiag}
C(t,t')&=&
+
\left\langle
\right\rangle + \ldots \nonumber\\
&=& 
+\ldots \nonumber\mathrm{,}\\
&\equiv& \int_0^t dt_1 \int_0^{t'}dt_2 G_0(t,t_1)D(t_1,t_2)G_0(t',t_2)\mathrm{,}
\end{eqnarray}
where to obtain the last line we defined
\begin{eqnarray}
D(t_1,t_2)\equiv2T\delta(t_1-t_2)+
\frac{g^2}{2}C_0(t_1,t_2)C_0(t_1,t_2)+\ldots
\end{eqnarray}

What about $G(t,t')$?
\begin{eqnarray}
G(t,t') =
 + \left\langle \frac{1}{2T}
\right\rangle  +\ldots\mathrm{,}
\end{eqnarray}
where the lone $\times$ has to be attached to
a $\times$ on the tree and then the remaining diagram is closed.
Thus,
\begin{eqnarray}\label{eq:Gdiag}
G(t,t') &=&
+\ldots\nonumber\mathrm{,}\\
&=&
G_0(t,t') + \int_{t'}^{t} dt_1 \int_{t'}^{t_1}dt_2
G_0(t,t_1)\Sigma(t_1,t_2)G_0(t_2,t')\mathrm{,}
\end{eqnarray}
where we similarly defined
\begin{eqnarray}
\Sigma (t_1,t_2) \equiv g^2 C_0(t_1,t_2) G_0(t_1,t_2)+\ldots
\end{eqnarray}
This is an exact,
formal representation of the perturbations series.  In fact, in
some sense it is simply a definition of the kernels $D$ and
$\Sigma$.  We can appeal to the structure of the perturbation
series to justify this.

Also, the lower limit of the second integration insures that
$t_2>t'$. In fact, let's take a closer at the limits of
integration for a sample diagram, the second term in Eq.~\ref{eq:Gdiag},
which is reproduced on the left-hand side of Fig.~\ref{fig:diagrams} with additional labels.
The incoming branch imposes $t>t_1$, the central loop $t_1>t_2$, and
the outgoing branch $t_2>t'$, for an overall $t>t_1>t_2>t'$. The
resulting integration limits are thus $\int_{t'}^{t} dt_1\int_{t'}^{t_1}dt_2$.

\begin{figure}
\begin{center}
\hspace{0.5in}
\end{center}
\caption{Left: As an exercise, work out the integration limits for this sample diagram.
The time pairings indicate the beginning and end times of a given segment.
There are in this diagram two internal vertices, $t_1$ and $t_2$, and two external ones, $t$ and $t'$.
Right: a simple tadpole diagram.}
\label{fig:diagrams}
\end{figure}

If you have followed so far, you might be asking
yourself what happened to the diagrams like the
one appearing on the right in Fig.~\ref{fig:diagrams}.
Such closed loops are called \emph{tadpoles}.
They do not contribute to the time dependence of $C$ or $G$, and
we assume that their contribution is absorbed into $\mu(t)$. The
next lowest-order terms are then the diagrammatic forms we last drew
in Eq.~\ref{eq:Cdiag} and \ref{eq:Gdiag}.

To make a self-consistent approximation we replace $G_0$ and $C_0$
in our second-order approximations for $D$ and $\Sigma$ by $G$ and
$C$.  We will call this the \emph{Mode-Coupling Approximation}, a
name that will be  clear in meaning at the end
\cite{bouchaud:1996,bouchaud:1997}. This is Eq.~\ref{eq:Cdiag} and
\ref{eq:Gdiag} with
\begin{eqnarray}
\Sigma(t_1,t_2) &=& g^2 C(t_1,t_2) G(t_1,t_2)\mathrm{,}\\
D(t_1,t_2) &=& 2T\delta(t_1,t_2) + \frac{g^2}{2}C(t_1,t_2)^2\mathrm{.}
\end{eqnarray}
These two equations can be further manipulated by noting that
\begin{eqnarray}
G_0=\left(\mu(t) + \frac{\partial}{\partial t}\right)^{-1}\mathrm{,}
\end{eqnarray}
and so we can multiply both sides of Eq.~\ref{eq:Cdiag} and
\ref{eq:Gdiag} by $G_0^{-1}$.
\begin{eqnarray}
G_0^{-1}\bigotimes G &=& I + \Sigma \bigotimes G\mathrm{,}\\
G_0^{-1}\bigotimes C
&=& G_0^{-1} \bigotimes \left\{G_0 + G_0\bigotimes\Sigma\bigotimes
G\right\} \bigotimes D\bigotimes G\nonumber\mathrm{,}\\
&=& D \bigotimes G + \Sigma \bigotimes
\underbrace{\left\{G \bigotimes D\bigotimes G
\right\}}_{C}\nonumber\mathrm{,}\\
&=& D\bigotimes G + \Sigma \bigotimes C\mathrm{,}
\end{eqnarray}
where $I$ is the identity operator. In other notation,

\noindent\fbox{
\begin{Beqnarray}
\left\{\frac{\partial}{\partial t} + \mu(t)\right\} G(t,t') &=&
\delta(t-t') + \int_{t'}^{t} dt'' \Sigma (t,t'') G(t'',t')\mathrm{,}\\
\left\{\frac{\partial}{\partial t} + \mu(t)\right\} C(t,t') &=&
\int_0^t dt'' D(t,t'') G(t',t'') + \int_0^t dt'' \Sigma(t,t'')
C(t'',t')\label{eq:Coperator}
\end{Beqnarray}}

This still does not look like the MCT equations derived
before. We will manipulate the RHS of the second equation to achieve this.
Taking the first term, we substitute
\begin{eqnarray}
\int_0^t dt'' D(t,t'')G(t',t'') \rightarrow \int_0^{t'}
D(t,t'')\frac{1}{T} \frac{\partial}{\partial t''} C(t',t'')\mathrm{.}
\end{eqnarray}
We can replace $t\rightarrow t'$ because of the restriction on
time arguments in $G(t',t'')$ and  we can substitute $G$ for $C$ as
written by using the \emph{fluctuation-dissipation theorem} (FDT) and
assuming that the system is at equilibrium. The result can now be integrated by parts,
\begin{eqnarray}
\frac{1}{T} \left.\left[D'(t,t'')C(t'',t')\right]\right|_0^{t'} -
\frac{1}{T} \int_0^{t'} dt'' \frac{\partial}{\partial
t''}D'(t,t'')C(t'',t')\mathrm{.}
\end{eqnarray}
Note that we can neglect the $\delta$-function part of $D$ here,
since it provides no contribution. So, we denote the regular part of $D$ as
$D'$.

Again, using the FDT, we can substitute
\begin{eqnarray}
\frac{\partial }{\partial t''}D'(t,t'') \rightarrow T\Sigma(t,t''),
\end{eqnarray}
yielding an integral which can be combined with the other term of
Eq.~\ref{eq:Coperator} to obtain
\begin{eqnarray}
-\int_0^{t'} dt'' \Sigma(t,t'')C(t'',t') + \int_0^{t} dt''
\Sigma(t,t'')C(t'',t')= \int_{t'}^{t} dt'' \Sigma(t,t'')C(t'',t')\mathrm{.}
\end{eqnarray}
Now, using the FDT in the reverse direction and integrating by parts, we get
\begin{eqnarray}
\frac{1}{T} \left.\left[D'(t,t'')C(t'',t')\right]\right|_{t'}^{t} -
\frac{1}{T} \int_{t'}^{t} dt'' D'(t,t'')\frac{\partial}{\partial
t''}C(t'',t')\mathrm{.}
\end{eqnarray}
Combining all terms on the right-hand side of
Eq.~\ref{eq:Coperator}, we get
\begin{eqnarray}
\frac{1}{T} \left[D'(t,t)C(t,t')-D'(t,0)C(0,t')\right] -
\frac{1}{T} \int_{t'}^{t} dt'' D'(t,t'')\frac{\partial}{\partial
t''}C(t'',t')\mathrm{.}
\end{eqnarray}
Using the fact that, in equilibrium, these functions are
time-translation invariant,
\begin{eqnarray}
D'(t,t'')&=& D'(t-t'')\nonumber\mathrm{,}\\
C(t,t')&=&C(t-t')\nonumber\mathrm{,}\\
C(t'',t') &=& C(t''-t')\nonumber\mathrm{,}\\
C(0,t')&=& C(t')\nonumber\mathrm{,}\\
D'(t,0)  &=& D'(t)\nonumber\mathrm{,}\\
D'(t,t)&=&D'(0)\mathrm{,}
\end{eqnarray}
and making the transformation,
\begin{eqnarray}
t''-t'&\equiv& \tau'\nonumber\mathrm{,}\\
t-t'&\equiv& \tau\nonumber\mathrm{,}\\
t'&\rightarrow& \infty\mathrm{,}
\end{eqnarray}
we find
\begin{eqnarray}
\left(\frac{\partial}{\partial t} + \tilde{\mu}(t)\right)
C(t)+\frac{1}{T}\int_0^t d\tau D'(t-\tau)\frac{\partial
C(\tau)}{\partial\tau} = 0\mathrm{,}
\end{eqnarray}
where
\begin{eqnarray}
\tilde{\mu}(t)&=&\mu(t)-\frac{1}{T}D'(0)\mathrm{,}\\
D'(t)&=&\frac{g^2}{2}C(t-\tau)^2\mathrm{.}
\end{eqnarray}

This is just the schematic model with
\begin{eqnarray}
\Omega_0^2 \leftrightarrow \tilde{\mu}(t)\mathrm{,}\\
\lambda  \leftrightarrow \frac{g^2}{2T}\mathrm{.}
\end{eqnarray}
The only difference is that $\frac{\partial C(t)}{\partial t}$
appears instead of $\frac{\partial^2 \Phi(t)}{\partial t^2}$. This
actually makes \emph{no} difference as far as the glassy
properties are concerned.  In fact, for models of overdamped
systems, such as Brownian colloidal spheres, $\frac{\partial
\Phi(t)}{\partial t}$ is what appears naturally in the reduction
of the full MCT equations of the schematic model.  This completes
the relationship between the memory function/projection operator
MCT derivation and a field-theoretic approach.

\section{Looking Ahead: Beyond ``Simple'' MCT}
In this section we will outline some thoughts on attempts to do better
than the MCT derived thus far.  This is not an exhaustive
discussion, but is meant to give an idea about what can
be done.

\subsection{Coupling to Currents}
G\"{o}tze and Sj\"{o}gren \cite{gotze:1987} as well as Das and
Mazenko \cite{das:1986} have developed theories that remove the
sharp transition at $T=T_0$. In both cases, it is the coupling to
certain current modes that are ignored in the expressions we have
just derived that restore ergodicity below $T_c$.

In the theory of G\"{o}tze and Sj\"{o}gren, the Laplace transform
of our exact equation of motion for $F(k,t)$ is given as
\begin{eqnarray}
F(k,z)=\frac{-1}{z-\frac{\Omega(k)^2}{z+M(k,z)}}\mathrm{,}
\end{eqnarray}
where $\Omega(k)=\frac{k^2 k_B T}{m S(k)}$ and
$M(k,z)$ is the Laplace transform of the memory function
\begin{eqnarray}
M(k,z)=\int_0^{\infty}dt e^{-zt}M(k,t)\mathrm{,}
\end{eqnarray}
and similarly,
\begin{eqnarray}
F(k,z)=\int_0^{\infty}dt e^{-zt}F(k,t)\mathrm{.}
\end{eqnarray}

Essentially, within the ``extended'' MCT of G\"{o}tze and Sj\"{o}gren,
\begin{eqnarray}
M(k,z)\sim\frac{K(k,z)}{1-\delta_h(k,z)K(k,z)}\mathrm{,}
\end{eqnarray}
where $K(k,z)$ is the ordinary MCT memory function given
in Eq.~\ref{eq:exactmemory}. The expression for
$\delta_h(k,z)$ is complicated, but to see what it does,
note that:
\begin{enumerate}
  \item if $\delta_h=0$, we recover exactly the ordinary MCT that we
  have derived before.  This can be checked by applying a Laplace
  transform the old expressions.
  \item if $\delta_h(k,z)$ has no singularities as
  $z\rightarrow0$, the the strict transition at $T_c$ in the MCT
  we have previously derived is removed since relaxation is
  governed by $M\sim1/|\delta_h|$ at long times and not
  $K(k,z)$, which yields a pole singularity in z-space.
\end{enumerate}

In the theory of Das and Mazenko, a hydrodynamic approach is used.
The kinetic energy of the free energy functional in terms of
current $j$ and density $\rho$ has the form
\begin{eqnarray}
K.E.[j,\rho] \sim\frac{j^2}{\rho}\mathrm{.}
\end{eqnarray}
This is like the usual $p^2/2m$ kinetic energy. However, the
$1/\rho$ part coupled to the current rounds off the strict
singularity at $T_c$, like in the G\"{o}tze and Sj\"{o}gren
theory.

Note that in both theories, we need currents to restore
ergodicity.  For some systems, like (simulated) colloidal hard
spheres undergoing Brownian motions, these currents do not exist!
Thus, the G\"{o}tze and Sj\"{o}gren and the Das and Mazenko
theories cannot be used to improve ordinary MCT there.

\subsection{New Closures}
As mentioned above, the extended MCT of G\"{o}tze and Sj\"{o}gren
and of Das and Mazenko cannot tell us anything (beyond ordinary MCT)
for Brownian hard-sphere systems.  An interesting proposal was
recently put forward by Szamel \cite{szamel:2003}.  The main idea is
\emph{not} to factorize the memory function expression leading to
the ordinary MCT given in Eq.~\ref{eq:exactmemory}, but to write an
\emph{exact} equation of motion for it, and then factorize the
memory function for the new equation.  Here is a sketch of the idea.

Recall our old approach to MCT,
\begin{eqnarray}
\frac{\partial^2F(k,t)}{\partial t^2}+\frac{k^2 k_BT}{m
S(k)}F(k,t)+\int_0^td\tau
K(k,t-\tau)\frac{\partial F(k,z)}{\partial\tau}=0\mathrm{,}
\end{eqnarray}
where essentially $K(k,t)\sim\langle
\delta\rho\delta\rho\delta\rho\delta\rho\rangle$ is a four-point function of density
variables. In the old approach, the closure involved
\begin{eqnarray}
K(k,t)\sim
\langle\delta\rho\delta\rho\rangle\langle\delta\rho\delta\rho\rangle
= \sum_{\mathbf{q}}F(q,t)F(|\mathbf{k}-\mathbf{q}|,t)
\end{eqnarray}
and this allowed to solve for $F(k,t)$.

Instead of factorizing the four-point memory kernel, let's write an
exact equation of motion for it, following the same lines of
reasoning as before,
\begin{eqnarray}
\frac{\partial^2K(t)}{\partial t^2} + \Gamma K(t) + \int_0^t d\tau
R(t-\tau)\frac{\partial K(\tau)}{\partial \tau}=0\mathrm{,}
\end{eqnarray}
The wavevector indices are suppressed to
simplify the notation in order to clarify idea behind the manipulations.
This has the same form as before, but with new frequencies
$\Gamma$ and a new memory function $R(t-\tau)$.

Schematically,
$R\sim\langle \delta\rho\delta\rho\delta\rho\delta\rho\delta\rho\delta\rho\rangle$
is a six-point function!  We can close the equation for $K$ and $F$, by approximating
$R\approx K\cdot F$ the product of a four-point and a two-point
function. This yields two coupled sets of integro-differential
equations that may be solved self-consistently yielding a converged
$F(k,t)$.

This approach has not been considered for the full dynamics of
$F(k,t)$, but yields a better estimate for $T_c$,
(\emph{i.e.} the $T_c$ that is extracted is closer to the measured
glass transition).

\subsection{Four-point correlations and dynamical heterogeneities}

\begin{figure}
  \includegraphics[width=\columnwidth]{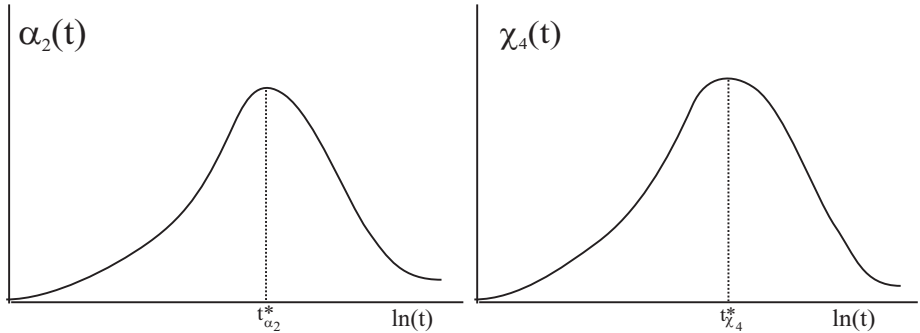}
  \caption{The non-Gaussian parameter $\alpha_2(t)$(left)
  and the four-point correlation function $\chi_4(t)$ (right)
  peak at times $t^*$ where $t_{\chi_4}^* > t_{\alpha_2}^*$.}\label{fig:tstar}
\end{figure}
It was mentioned in Section \ref{sec:MCT} that MCT does not describe well
\begin{eqnarray}
\alpha_2(t)\equiv\frac{3\langle
r^4(t)\rangle}{5\langle r^2(t)\rangle^2}-1
\end{eqnarray}
and that $\alpha_2(t)$ seems to correlate well with the timescale of
maximal dynamical heterogeneity \cite{donati:1999,kob:1997}. It
turns out that this timescale is in the late $\beta$-regime. This
highlights the fact $\alpha_2(t)$ yields information on  transiently
mobile particles that jump due to the destruction of cages.  It
should be noted that there is no lengthscale dependence in
$\alpha_2(t)$.

To gain some information about a growing (dynamical) lengthscale, a
multipoint dynamical generalization of the static structure factor
may be studied
\cite{glotzer:2000,biroli:2004,franz:2000,kirkpatrick:1988}. The
$k\rightarrow 0$ limit of this structure factor, as Sharon Glotzer
discusses in her lectures, is the susceptibility
\begin{eqnarray}
\chi_4(t)&\sim& \int d\mathbf{r}_1 \ldots d\mathbf{r}_4
\theta_a\left(|\mathbf{r}_1-\mathbf{r}_2|\right)
\theta_a\left(|\mathbf{r}_3-\mathbf{r}_4|\right)\times\nonumber\\
& &\left\langle \rho(\mathbf{r}_1,0) \rho(\mathbf{r}_2,t)
\rho(\mathbf{r}_3,0) \rho(\mathbf{r}_4,t)\right \rangle\mathrm{,}
\end{eqnarray}
where the function
$\theta_a\left(|\mathbf{r}_1-\mathbf{r}_2|\right)$ equals one when
$|\mathbf{r}_1-\mathbf{r}_2|\leq a$, and zero otherwise
\cite{glotzer:2000}. The timescale at which $\chi_4(t)$ peaks is
generally in the $\alpha$-regime. The growing lengthscale associated
with dynamic heterogeneity is associated with slow moving,
transiently caged particles.

Given the superficial similarity with $\alpha_2(t)$, as shown in
figure\ref{fig:tstar}, one might conclude that MCT cannot compute
objects like $\chi_4$, but this is not the case. Recent work by
Biroli and Bouchaud \cite{biroli:2004}, motivated by the earlier
insight of Franz and Parisi \cite{franz:2000} and Kirkpatrick and
Thirumalai \cite{kirkpatrick:1988}, shows that MCT may make
quantitative statements about the scales of length and time
associated with dynamical heterogeneity.  So far, absolute
lengthscales have not been computed, but dynamical exponents $z$
relating timescales $\tau$ and lengthscales $\xi$ have,
\begin{eqnarray}
\tau \sim \xi^z\mathrm{,}
\end{eqnarray}
where $z=2\gamma$ and $\gamma$ is given in Eq.~\ref{eq:gammaexp}.

\bibliographystyle{unsrt}
\bibliography{IndiaNotes}

\end{document}